# Activated Dissociation of $H_2$ on Cu(001): The Role of Quantum Tunneling

Xiaofan Yu[1,2,#] (于小凡), Yangwu Tong[1,2,#] (童洋武), and Yong Yang[1,2,*] (杨勇)

1. *Key Lab of Photovoltaic and Energy Conservation Materials, Institute of Solid State Physics, HFIPS, Chinese Academy of Sciences, Hefei 230031, China.*
2. *Science Island Branch of Graduate School, University of Science and Technology of China, Hefei 230026, China.*

**ABSTRACT:**

The activation and dissociation of $H_2$ molecules on Cu(001) surface is studied theoretically. The activation barrier for the dissociation of $H_2$ on Cu(001) is determined by first-principles calculations to be ~ 0.59 eV in height. Electron transfer from the substrate Cu to $H_2$ plays a key role in the activation, breaking of the H-H bond and the formation of the Cu-H bonds. At around the critical height of bond breaking, two stationary states are identified, which correspond respectively to the molecular and dissociative state. Using the transfer matrix method, we are able to study the role of quantum tunneling in the dissociation process along the minimum energy pathway (MEP), which is found to be significant at room temperature and below. At given temperatures, the tunneling contributions from the translational and vibrational motions of $H_2$ are quantified for the dissociation process. Within a wide range of temperatures, the effects of quantum tunneling on the effective barriers of dissociation and the rate constants are revealed. The deduced energetic parameters associated with thermal equilibrium and non-equilibrium (molecular beam) conditions are comparable with experimental data. In the low-temperature region, crossover from classical to quantum regime is identified.

[#]These authors contribute equally to this work.

[*]Corresponding author: yyanglab@issp.ac.cn

# 1. Introduction

As a clean energy source, hydrogen mainly exists in the form of molecules in nature. Materials involving metal elements are commonly utilized as catalysts for hydrogen production and storage to harvesting hydrogen based energy [1-3]. The elementary dynamics (e.g., adsorption and diffusion) of hydrogen on transition metal surfaces are closely related to some important physical and chemical processes such as crystal growth, hydrogen embrittlement, surface corrosion, and technological applications like radiation protection in fusion reactor reactions, electrode reactions in fuel cells, and surface catalysis [4-19]. When $H_2$ molecules approach a metal surface, depending on the energetics, they may exist in the molecular form, or the dissociative form with the H atoms attached separately on the metal surface. The key factor determining the energetics is the energy path in particular the energy barrier associated with the dissociation process, which is referred to as the dissociation barrier hereafter. It is found that the dissociation barriers of hydrogen molecules on different metal surfaces are different. For instance, on the surfaces of Pt and Rh, the adsorbed $H_2$ molecules will spontaneously decompose into H atoms [20-22]. By contrast, on the surface of Cu and Ag [23], the adsorbed $H_2$ molecules do not spontaneously decompose into H atoms, and the dissociation process requires additional energy consumption, which is usually referred to as an activated process. In the spontaneous process, the related energy barrier $E_b = 0$ whereas $E_b > 0$ in the activated process. The dissociation barriers of hydrogen on different crystallographic planes are also quite different. The existence of an activation barrier is commonly believed to be due to the full occupation of $d$ orbitals which induces Pauli repulsion when molecules and surfaces come into contact [24, 25]. Consequently, dissociation of $H_2$ on the metal surface with fully occupied $d$ orbitals tends to be an activated process ($E_b > 0$), while dissociation on the metal surface with partially occupied $d$ orbitals is more likely to be a spontaneous process ($E_b = 0$). There have been many experimental reports [6, 8, 10, 14, 15] on the adsorption, dissociation and diffusion processes of hydrogen on different metal surfaces. For example, nozzle beam experiment in combination with sticking coefficient measurements for monoenergetic molecular beams as a function of energy and angle of incidence were carried out to study the adsorption/desorption kinetics of $H_2$ on Cu surfaces [6]. Using thermal desorption in combination with work function and low electron diffraction (LEED) measurements, the H/Pd(100) system has been studied to determine the adsorption state and adsorption energy [8]. Molecular beam techniques were utilized to control the incident kinetic energy and incident angle of $H_2$ molecules to explore in detail the nature of the dissociation barriers on Cu [10].

As a typical prototype system for the activated reactions of $H_2$ with metal surfaces, $H_2$/Cu has been the subject of immense theoretical investigations, focusing primarily on the dissociation and sticking dynamics [26-49], and the thermal desorption spectra [50], which are directly related to the energy pathway governing the dissociation/desorption processes. Generally, the collision of a diatomic molecule with surface involves at least six degrees of freedom when the surface is approximately taken as rigid. For accurate description of the process, six-dimensional (6D) dynamics study is usually required, which is feasible for classical mechanics while relatively challenging for quantum level simulations. Based on the potential energy surface (PES) constructed within the Born-Oppenheimer approximation, multi-dimensional (3, 4, 5, 6D) quantum dynamics calculations have been carried out to study the activated dissociation of $H_2$ on Cu surfaces [28-41], with focus on the role of vibrational and rotational states of $H_2$, as well as the coupling of both degrees of freedom [29-40]. It is found that by 6D quantum simulations that [33-35] the dissociation probability depends on the orientation of rotation of the incident $H_2$ molecule: The so-called ''helicopter'' orientation (The angular momentum vectors oriented perpendicular to the surface, or the line joining the H-H bond is parallel to the surface) yields dramatically larger reaction probability than the ''cartwheel'' orientation, in which the angular momentum vectors of $H_2$ oriented parallel to the surface. In recent years, special attention is paid to the development of chemically accurate PES for the interactions between $H_2$ and Cu and other metal surfaces [42-49]. Notable progress in this field includes, for instance, the implementation of semi-empirical specific reaction parameter (SRP) approach to density functional theory (DFT) [42-47], and the recently developed machine learning PES [48, 49].

Compared to the other elements, hydrogen has a smaller mass and therefore significant nuclear quantum effects are expected in its dynamical process [51-54]. In recent years, the important role of atomic quantum tunneling plays in the reaction rates and selectivity of chemical processes involving small molecules has been increasingly reported [55, 56]. For instance, it is found that nuclear quantum effects account for the facile activation and dissociation of $H_2$ on Cu(111) surface which is alloyed with 1% monolayer of Pd [57, 58]. Analysis based on model potentials has shown that quantum tunneling is essential to the sticking of $H_2$ on metal surfaces [26]. Despite the enormous number of experimental and theoretical investigations, there still lacks systematic and in-depth studies on the role of quantum tunneling in the activated dissociation of $H_2$ on metal surfaces. In the present work, we revisit this topic by studying the activation and dissociation of $H_2$ molecules on Cu(001). The effects of quantum tunneling are investigated based on first-principles calculations combined with the transfer matrix method, for which the minimum energy pathway (MEP) of dissociation is extracted from *ab initio* potential energy surface (PES). During the activation process of $H_2$ on Cu(001), it is shown that charge transfer between the Cu surface and the $H_2$ molecules is the key for the breaking of H-H bond and the formation of Cu-H bonds. Bistability of the adsorption states is identified at the vicinity of the critical point of dissociation. The probability of dissociation and the corresponding rate constants due to the translational ($H_2$ tunneling as a whole unit) and vibrational motions of $H_2$ molecules are quantitatively evaluated. The obtained activation barrier and the threshold kinetic energy for detectable dissociation events are in agreement with experimental observations. The effective dissociation barrier is evaluated as a function of temperature, whose magnitude is found to be significantly reduced owing to the quantum tunneling of H atoms. For the dissociation of hydrogen isotopes $H_2$ and $D_2$, the crossover from the high-temperature classical dynamics to low-temperature quantum dynamics is recognized.

This paper is organized as follows. Following the introductory part, Section 2 presents the technical details of first-principles calculations and the transfer matrix (TM) method employed in this study. In Section 3, the results on the energy path of the activated dissociation of $H_2$ on Cu(001), the role of charge transfer and the role of quantum tunneling in shaping the kinetic process are presented and compared with experiments. Section 4 summarizes the main conclusions.

## 2. Methods
### 2.1 Details of First-principles Calculations

The Vienna *ab initio* simulation package (VASP) [59, 60] based on density functional theory (DFT) is employed for the first-principles calculations. The Perdew-Burke-Ernzerhof (PBE) type functional within the generalized gradient approximation (GGA) [61, 62] is used to describe the exchange-correlation terms of electrons, in combination with the PAW potentials [63, 64] to describe the electron-ion interactions. The energy cutoff for plane wave basis sets is 600 eV. The initial atomic configurations are constructed with the aid of VESTA [65], in which the Cu(001) surface is modeled by a six-layer $p(3\times3)$ supercell, repeating periodically along the *xy* plane with a vacuum layer of about 15 Å along the *z* direction. In all the calculations, the Cu atoms in the bottom three layers are fixed and the atoms in the upper layers are relaxed. We employ a dipole correction for the total energy to eliminate the artificial dipole-dipole interaction caused by the upper and lower asymmetric slab surfaces. A 4×4×1 Monkhorst-Pack k-mesh [66] is generated for sampling the Brillouin zone (BZ) in performing structural relaxation and total energy calculations. These set of parameters ensure the total energy calculations to converge within an error bar of ~ 0.5 meV/atom. The vibrational properties of the relaxed structures are analyzed using density functional perturbation theory (DFPT) [67].

### 2.2 The Transfer Matrix Method

The probability of a quantum particle passing through a potential barrier of arbitrary shape can be obtained using the transfer matrix (TM) method. By numerically slicing an arbitrarily shaped potential, it is transformed into a stack of multiple rectangular potential barriers (potential wells) [54]. The transmission of a particle through each rectangular potential barrier (potential well) can be represented by a matrix of coefficients that describe the transmitted and reflected amplitudes of the wave function. Multiplying the coefficient matrices in turn one is able to obtain a transition matrix representing the transition relationship between the initial and final state. For the transmission across a potential $V(x)$, the incoming and outgoing amplitudes ($A_L$, $B_L$; $A_R$, $B_R$) of the wave functions can be related to each other as follows [54]:

$$\begin{pmatrix} A_R \\ B_R \end{pmatrix} = M \begin{pmatrix} A_L \\ B_L \end{pmatrix} \quad (1)$$

where $M \equiv \begin{pmatrix} m_{11} & m_{12} \\ m_{21} & m_{22} \end{pmatrix}$ is the transfer matrix. In a system which preserves the time-reversal symmetry, the determinant $|M| = 1$, and the transmission coefficient is calculated by $T_r(E) = \frac{1}{|m_{22}|^2}$.

The advantage of the TM method is that the obtained transmission probabilities are numerically accurate with comparison to that calculated by the Wentzel–Kramers–Brillouin (WKB) approximation [68]. For instance, when the incident energy of a classical particle is higher than the potential barrier, the probability of the particle passing through the barrier is 1. Such a deficiency of traditional WKB is partly amended by its impoved version [69]. For a quantum particle, due to the existence of quantum interference, even if the energy of the particle is greater than the barrier height, there may be a certain probability of reflection which makes the probability of passing through the potential barrier smaller than 1. The TM method deals with the transport of quantum particles through a given potential field in a unified manner, and fully takes into account quantum effects in the process of crossing a potential barrier.

## 3. Results & Discussion

### 3.1. Mechanistic analysis of the interaction between $H_2$ and Cu(001)

We begin by investigating the atomistic process of $H_2$ dissociation and adsorption on the Cu(001) surface. Shown in Fig. 1, are some typical configurations representing the dissociation process of a $H_2$ molecule which approaches the Cu(001) along the surface normal ($z$) direction. The orientation line of H-H bond is parallel to Cu(001), the midpoint of H-H bond is right above one of the bridge sites of Cu(001) with the projection of H-H bond perpendicular to the surface Cu-Cu bond. Such an initial configuration gives rise to the largest possibility of dissociation [17, 25, 33-35]. It is found by our calculations that, when the distance between the $H_2$ molecule and the Cu (001) surface decreases to a critical value ($z_c$ = 0.947 Å), the covalent bond in the $H_2$ molecule tends to get broken and decomposes into two adsorbed H atoms on the Cu(001). The corresponding PES is obtained by the following procedure: At a height of center of mass $Z_{H2}$, the $H_2$ bond length $d_{H-H}$ is gradually varied, and the configuration space is sampled point-by-point with the energy of each configuration at the coordinates ($Z_{H2}$, $d_{H-H}$) given by DFT calculations. The PES is shown in Fig. 2(a), with its 2D projection onto the parameter plane $Z_{H2}$-$d_{H-H}$ shown as a contour plot in Fig. 2(b). The MEP is indicated by the scattered dots on Fig. 2(a) and 2(b) with four typical configurations labeled by capital letters A, B, C, and D, whose top and side views are schematically shown in Fig. 1. The relative energy with referenced to the initial molecular state (configuration A) is displayed in Fig. 2(c), as a function of the distance ($|Z_0$-$Z|$) travelled by the center of mass of $H_2$ when approaching the Cu(001) surface. Compared to Figs. 2(a)-(b), the relative energies for the configurations near the critical point shown in Fig. 2(c) are refined by calculations with relaxed $H_2$ bond length ($d_{H-H}$) at given height $Z_{H2}$. Therefore, the refined MEP represents a set of energy local minima along the reaction path. Figure 2(d) shows the variation of total energy $E_0$ (with referenced to configuration A) as a function of $H_2$ bond length $d_{H-H}$ at a series of heights $z$ (= $Z_{H2}$), from the gaseous molecular state ($z \sim 2.813$ Å) to near the critical height of dissociation ($z_c \sim 0.947$Å). The local minima of the curves correspond to the dynamically stable configurations locating on the MEP, which yield the stable bond lengths of $H_2$ $d_{H-H}$ at a given height. Compared to the other $E_0$-$d_{H-H}$ curves of molecular state, the most remarkable feature of the curve ($z$ = 0.968 Å) at the vicinity of the critical height ($z_c$ = 0.947 Å) is existence of two stationary points as indicated by the two local minima at $d_{H-H} \sim 0.990$ Å and $\sim 2.772$ Å, respectively, which are separated by a small barrier of $\Delta E \sim 3$ meV. The small barrier implies that a weak perturbation will switch the molecular state (C′) to the dissociative state (C′$_d$), leading to an abrupt change of the H$_2$ bond length d$_{H-H}$. At temperatures $T \leq 300$ K, the H$_2$ molecule is frozen at its vibrational ground state, the translational motion along the surface normal direction ($z$) therefore plays a major role. Near the critical height, a slight increase of kinetic energy of the center of mass and consequently a small decrease of the H$_2$-Cu(001) distance will result in abrupt breaking of the H-H bond.

Before the onset of dissociation, energy of the system is continuously rising with decreasing H$_2$-Cu(001) distance. At the critical point of bond breaking, there is a sudden change and the energy of stationary states drops abruptly, as indicated by the C-C$_d$ line shown in Fig. 2(c). Accordingly, the energy drop (from configuration C to D) in this process corresponds to the desorption barrier of chemically adsorbed H on Cu(001). The height of the barrier ($E_b \sim 0.586$ eV) is in good agreement with previous works ($E_b \sim 0.58$ eV) [17, 46, 70] at the GGA level and is lower than the barrier height ($E_b \sim 0.74$ eV) calculated using the SRP-DFT method [43, 45, 46]. Experimentally, temperature programmed desorption (TPD) is commonly employed to measure the desorption barrier and desorption rate of H$_2$ on the Cu(001) surface [15], which can be numerically simulated based on DFT calculations in combination with kinetics analysis [50].

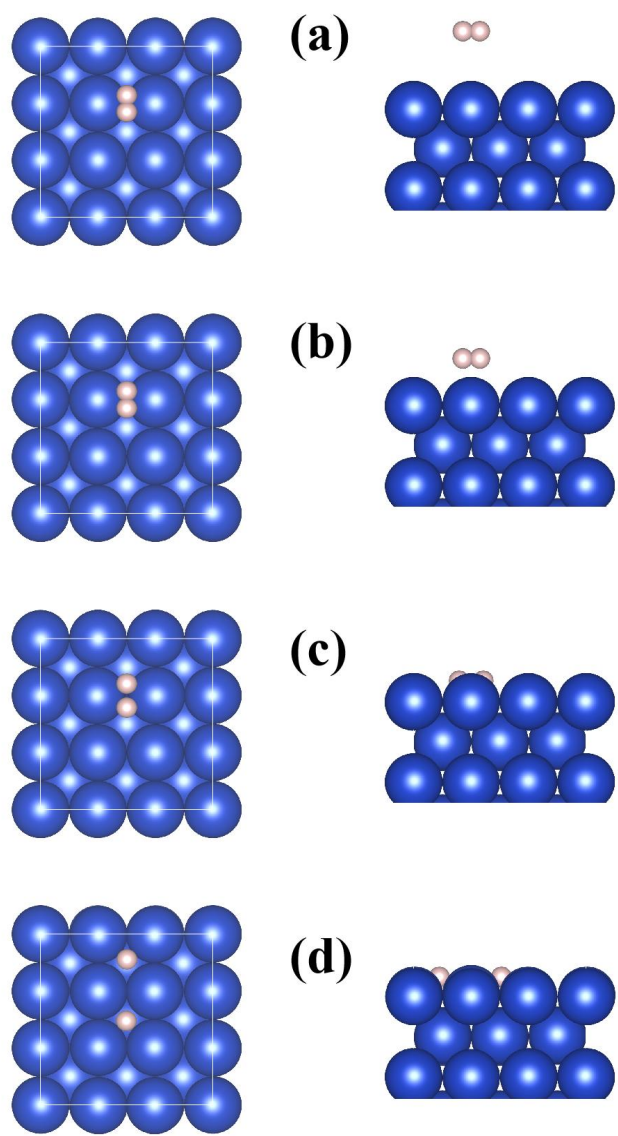

**Fig. 1**. Top (left) and side views (right) of typical $H_2$/Cu (001) configurations from the molecular state to the dissociative state. The blue (large) balls represent for Cu atoms while the pink (small) ones for H atoms. This convention applies to all the figures.

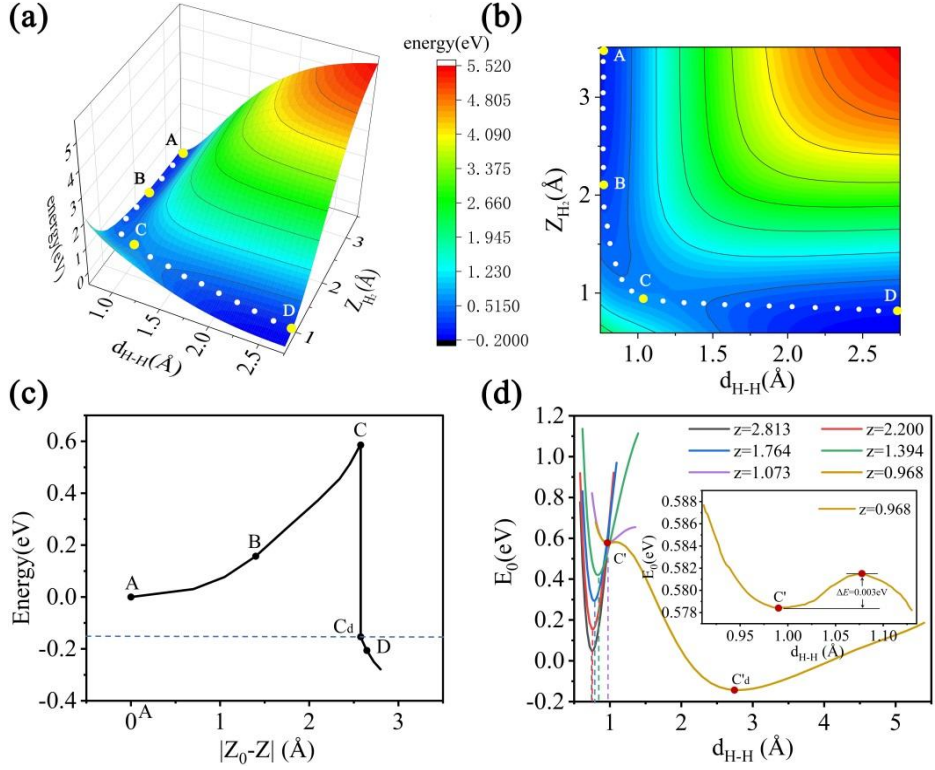

**Fig. 2.** Calculated PES (a), its 2D contours (b), and the refined MEP (c) for the activated dissociation of $H_2$ on Cu(001); (d) variation of relative energy ($E_0$) with respect to the $H_2$ bond lengths ($d_{H-H}$) at given height $z$. The energy of initial configuration at gaseous phase (Configuration A) is set as zero point. The capital letters A-D have a one-to-one correspondence to the atomic configurations displayed in Figs. 1(a)-1(d). $Z_0$ is the initial height of $H_2$ and $|Z_0-Z|$ is the distance travelled. The capital letters C and $C_d$ represent the two stationary states at the critical height $z_c$.

To understand the bond breaking from the level of electron, we have calculated the charge density differences for a number of typical adsorption configurations, including the initial molecular state, the intermediate and transition states of dissociation, and the final fully dissociated state on Cu(001). Practically, the charge density difference ($\Delta\rho$) of a given configuration is obtained by subtracting the charge density of substrate and two individual H atoms from the total charge density of the adsorption system. The formula is as follows:

$$\Delta\rho = \rho[H_2/Cu(001)] - \rho[H\_1] - \rho[H\_2] - \rho[Cu(001)], \qquad (2)$$

where $\rho[H_2/Cu(001)]$ is the charge density of the system of $H_2/Cu(001)$, $\rho[H\_1]$

and $\rho[H\_2]$ denotes respectively the charge density of the two H atoms, $\rho[H_2/Cu(001)]$ is the total charge density of the adsorption system.

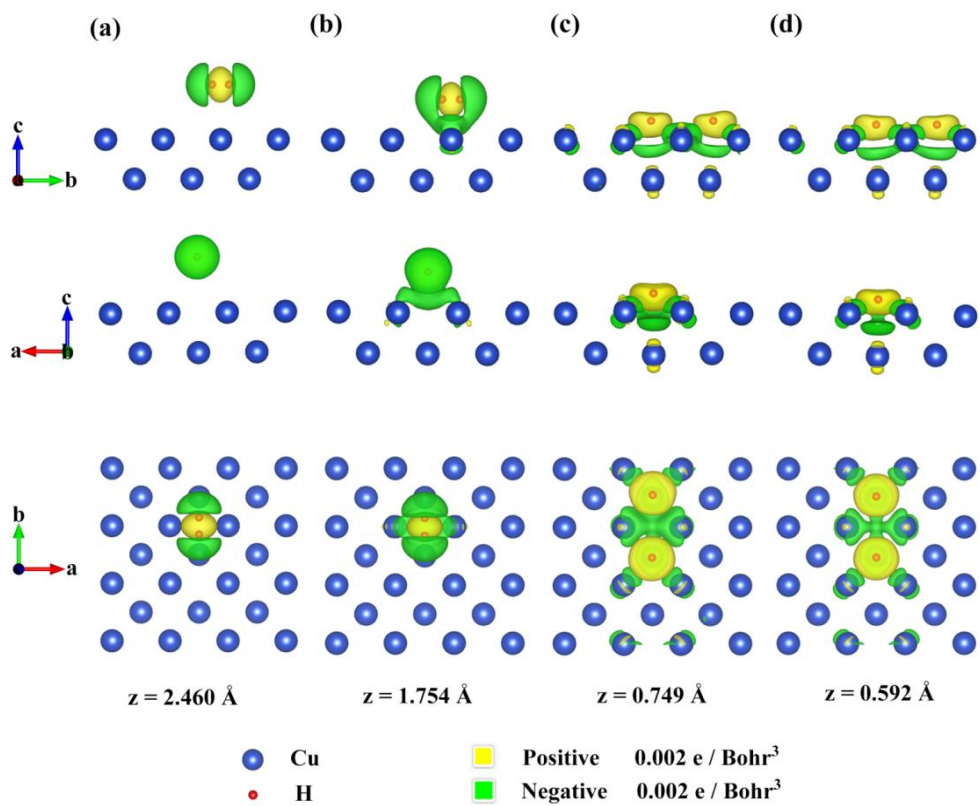

**Fig. 3**. Isosurfaces of charge density difference for the $H_2/Cu(001)$ at gradually decreasing $H_2$-Cu(001) distances. **(a)** The initial configuration of $H_2$, corresponding to the configuration in Fig. 1(a) and point A in Fig. 2. **(b)** The transition state of $H_2$ from molecular state to dissociation state. **Panels (c) and (d)**: The dissociative states in the form of adsorbed H atoms on Cu(001), with panel (d) being the final stable dissociative state.

The isosurfaces of charge density differences are shown in Fig. 3. When the $H_2$ molecule is far away from the Cu(001) surface, there is no charge transfer between $H_2$ and the substrate. Instead, one sees mainly the charge difference with positive values in-between the two H atoms (Fig. 3(a)), a clear evidence of the H-H covalent bond. As the distance between $H_2$ and Cu(001) decreases, the electrons from substrate Cu start to transfer to the H atoms. The closer they are, the larger amount of charge transfer is found. When the $H_2$-Cu(001) distance is $z = 0.749$ Å, the two H atoms are separated from each other and the H-H covalent bond is broken. The H atoms are located between four nearest neighboring Cu atoms and exchange charges with them. The H atoms gain electrons and the Cu atoms lose electrons. The electrons gather around the H atoms and indicate some characteristics of ionic bond between H and Cu atoms. Figure 4(a-d) shows the two-dimensional (2D) contours of charge distribution, which is sliced along the surface normal plane through the centers of the two H atoms. When the $H_2$-Cu(001) distance $z$ = 0.592 Å, the covalent bond is completely broken. It is worth noting that the H-H bond is very weak at $z_c$ = 0.947 Å, but it is still not broken, which can be regarded as the critical state. The breaking of the H-H bond is observed to be instantaneous when $H_2$-Cu(001) distance exceeds the critical value ($z_c$ = 0.947 Å).

To get more insights into the dynamical process associated with charge transfer, we study the charge transfer between $H_2$ molecule and the substrate using the Bader analysis [71]. The results are shown in Fig. 4(e). When a $H_2$ molecule gradually approaches the Cu(001), electrons from the underlying Cu substrate are gradually transferred to the two H atoms. The critical distance between the hydrogen molecule and the surface of Cu(001) is $Z_{H2-Cu(001)}$ ≈ 0.947 Å, corresponding to a decrease of height $|Z_0-Z|$ ≈ 2.576 Å. As seen from Fig. 2(c), the total energy of the system drops steeply at this position. Meanwhile, a sharp increasing of charge transfer to H atoms is observed (Fig. 4(e)). The abrupt change in both the total energy and the number of transferred electrons indicates that the dissociation of hydrogen molecules at the critical point may be an ultrafast process.

Figures 4(f-i) show the variation of 2D charge distribution between the one of the H (the other one is geometrically equivalent) and the nearest neighboring Cu atoms. At the initial distance $z$ = 2.460 Å, the H-Cu bond does not exist. As the distance between the $H_2$ molecule and the Cu(001) surface decreases, the H-Cu bond is gradually formed and strengthened. Finally, the H atoms form H-Cu bonds with the nearest neighboring Cu atoms. The charge contours clearly demonstrate the formation of H-Cu bonds when a $H_2$ molecule approaches the Cu(001) surface.

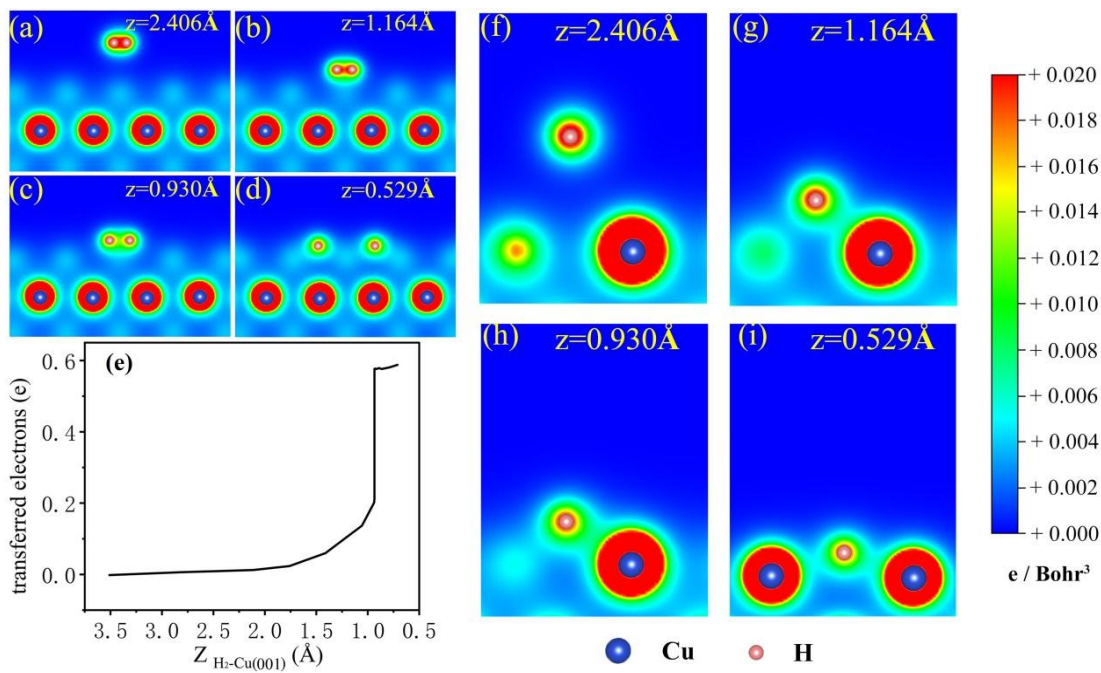

**Fig. 4**. (a-d): 2D contours of charge densities associated with the breaking of H-H bond, from the molecular state (a), the transition states (b, c), to the dissociative state (d). (e) Charge transfer from Cu(001) to $H_2$ based on Bader analysis, as a function of the distance from the center of mass of $H_2$ to Cu(001). (f-i): 2D contours of charge densities illustrating the formation of H-Cu bond, for the same configurations as panels (a-d) but viewed from different perspective angles.

From the analysis above, one may come to realize that the adsorption and dissociation of $H_2$ molecules on Cu(001) is essentially the competition between H-H and H-Cu bonds. Regarding the forces governing this process, the nuclear (ion core) – electron interactions are attractive, and the nuclear (ion core) – nuclear (ion core), electron – electron interactions are repulsive. The balance between the attractive and repulsive forces leads to the formation of chemical bonds. The strength of attraction is largely related to the extent of overlap of electron wave functions. As a result, the strength of attraction and bonding between two atoms are largely determined by the overlap of wave functions of electrons that participate in the bond formation. For a given atom in molecular form, when the attraction from a foreign species of atom is larger than the one it bonded with, breaking of the old bond and formation of a new bond is therefore expected.

**3.2. Role of quantum tunneling in the dissociation processes**

For a $H_2$ molecule incident on the Cu(001), the line joining the H-H bond can be parallel or perpendicular to Cu(001). Generally, the orientation of H-H bonds with respect to the Cu(001) surface can be described by an angle α, which belongs to the range of $0 \leq α \leq 90°$. Previous studies based on semi-empirical as well as *ab initio* methods have shown that the barrier for $H_2$ dissociation on Cu(100) depends on the orientation angle [17, 33-35]. Dissociation occurs more easily when the H-H axis is parallel to the surface. This is reasonable since when the $H_2$ molecule is close to the Cu(001), it has the largest contact with the surface and therefore the biggest chance of reaction and dissociation. Therefore, we only consider the case of α = 0 here, for which the axis of the H-H bond is parallel to Cu(001) with the two H atoms and the center of mass sharing the same *z* coordinates. From the dynamical point of view, the dissociation of $H_2$ molecules on Cu(001) is mainly due to the contribution of two processes. The first process is the dissociation due to the translational motion of $H_2$ in the vertical direction along the MEP, the most probable reaction path which is determined with the variation of H-H bond length ($d_{H-H}$) intrinsically included in our DFT calculations. As shown in Fig. 5(a), a $H_2$ molecule approaches Cu(001) along the *z* direction by overcoming the potential barrier $U(z)$, i.e., the MEP, and finally the $H_2$ molecule reaches the state of dissociation and chemisorption. The second process is the dissociation due to the stretching vibrations of H atoms in the horizontal direction (parallel to Cu(001)). These two processes are respectively analogous to the concept of overcoming the early and late barriers in the dissociative attachment of diatomic molecules on surfaces [72]. As shown in Fig. 5(b), while approaching the Cu(001) along *z* direction, the lateral vibrations of each H atom provides an additional probability of breaking the H-H bond by overcoming the lateral potential barrier $U(x, y)$. This can be determined by standard DFT calculations. For a given temperature *T*, the dissociation probabilities of the two processes can be calculated separately, and the total dissociation probability is obtained by summing up the probabilities of the two processes. Unless otherwise stated, both $H_2$ molecules and single H atoms are treated as quantum particles to study the role of tunneling.

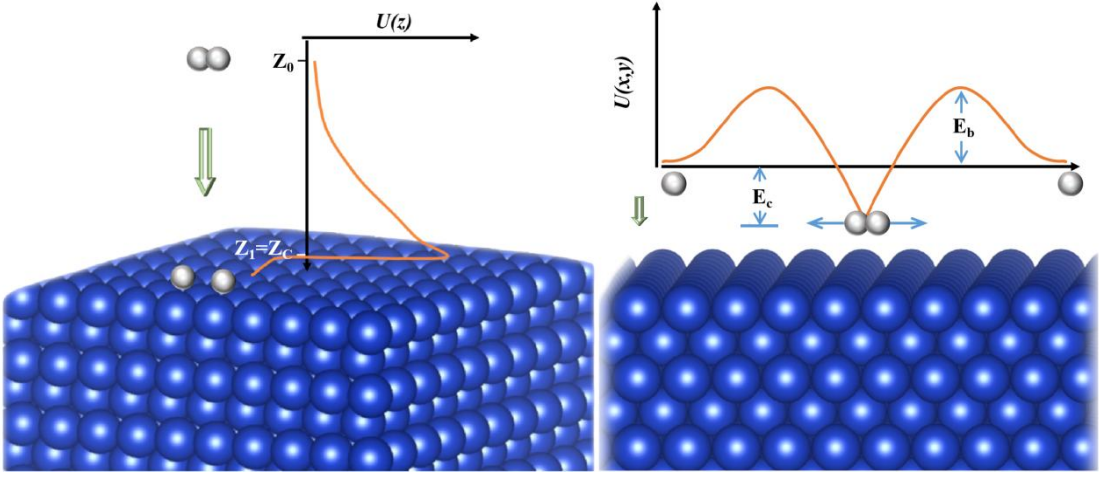

**Fig. 5**. Schematic diagrams for **(a)** Dissociation of $H_2$ molecules incident along the surface normal (*z*) directions. $Z_C$ and $Z_0$ correspond to the initial position of $H_2$ and the critical position (point C in Fig. 2). **(b)** The dissociation of $H_2$ molecules along the transverse direction. The potential energy difference between the instantaneous ground-state energy of $H_2$ at height *z* and the potential energy zero point (completely dissociative state) is marked as $E_c$. The potential barrier height to be overcome with reference to potential energy zero is marked as $E_b$.

For the first process, the probability of dissociation is

$$P_\perp(T) = \int_0^\infty p(E,T) T_r(E) dE \cong \int_0^{E_m} p(E,T) T_r(E) dE, \qquad (3)$$

where $p(E,T) = 2\pi(\frac{1}{\pi k_B T})^{3/2}\sqrt{E}e^{-E/k_B T}$ is the kinetic energy distribution function [73], which is suitable for the particles in thermal equilibrium systems where parabolic momentum-energy relation presents and scalar potentials dominate the interactions [54], $T$ is the absolute temperature, $E$ is the incident energy, $k_B$ is the Boltzmann constant. The term $T_r(E)$ is the transmission probability of the $H_2$ molecule as whole unit across the potential barrier $U(z)$ (Fig. 5(a)) at a certain energy $E$, which is calculated by transfer matrix (TM) method. In practice, an upper bound ($E_m$) is set for evaluation of the integral, which ensures the numerical results to converge to desired precision.

For the second process, the probability of dissociation is

$$P_{//}(T) = \frac{1}{L}\int_{z_0}^{z_1} dz \int_{U(z)}^{\infty} p(E,T)T_r(z)dE \cong \frac{1}{L}\int_{z_0}^{z_1} dz \int_{U(z)}^{E_m} p(E,T)T_r(z)dE, \qquad (4)$$

where $z_0$ is the initial height of $H_2$ and $z_1$ is the minimum height for $H_2$ to remain at molecular state (equals to the critical height, $z_c$), as shown in Fig. 5(a). The length $L = (z_1 - z_0)$ is the normalization factor, which is the decrease of height with reference to Cu(001). The integration of the kinetic energy function is considered only for the case when $E \geq U(z)$. The term $T_r(z)$ is the probability that a $H_2$ molecule passes through the transversal potential $U(x, y)$ (Fig. 5(b)) at temperature $T$ and height $z$, which can be obtained by the transfer matrix (TM) method combined with calculations based on statistical mechanics. Quantum tunneling is possible only when the vibrational energy $E_n \geq E_c$, where $E_c$ is the energy difference between the molecular and dissociated state at a height $z$ (Fig. 5(b)). The mathematical expression for transmission probability is

$$T_r(z) = \sum_{n=n_c}^{\infty} p_n \times T_r(E_n - E_c, z), \qquad (5)$$

where $p_n = \frac{e^{-\beta E_n}}{Q}$, $\beta = \frac{1}{k_B T}$, $E_n = (n + \frac{1}{2})\hbar\omega$, $Q = \sum_{n=0}^{\infty} e^{-\beta E_n} = \frac{e^{\beta\hbar\omega/2}}{e^{\beta\hbar\omega}-1}$, $E_n$ is the $n$th vibrational energy level of $H_2$ within the harmonic approximation (frequency $\omega$), and the value of $\hbar\omega$ can be obtained directly from DFPT calculations. Using the above expressions, the probability $p_n$ can be expressed as

$$p_n = e^{-n\beta\hbar\omega}(1 - e^{-\beta\hbar\omega}). \qquad (6)$$

The term $T_r(E_n - E_c, z)$ is the probability of a $H_2$ molecule passing through the transversal barrier $U(x, y)$ with the incident energy of $E_n - E_c$ when the $H_2$ molecule is at the height $z$. Similarly, the transmission probability can be calculated using the TM method. The barrier height is $E_b$, as depicted in Fig. 5(b). For numerical evaluation with energy sampling interval in the order of magnitude of 0.1 meV, good convergence (a relative deviation of less than 0.1%) of the integral is obtained when $E_m \approx 5E_b$.

At a given height $z$, the energy of the system is set to the zero point of potential energy when the H-H bond is broken (Fig. 5(b)). The difference between the lowest potential energy point at this height and the potential energy zero is labeled as $E_c$. It should be emphasized that only the energy levels with $E_n \geq E_c$ can produce effective quantum tunneling, and the corresponding vibrational energy level is labeled as the $n_c$th excited state. The subscript of the summation is therefore $n \geq n_c$. In the case of $E_n < E_c$, evanescent waves of incident particles are expected, which decay exponentially with the distance travelled. Finally, at temperature $T$, the total probability of H$_2$ dissociation on Cu(001) as a quantum particle is given by $P_Q(T) = P_\perp(T) + P_{//}(T)$.

The role of quantum tunneling can be demonstrated by assuming the H$_2$ molecule to be a classical particle and then compare the probability of dissociation at given temperatures. Similarly, dissociation can be considered as due to the contributions from two processes. The first process is to overcome the potential barrier $U(z)$ to reach the decomposed adsorption state when H$_2$ gradually approaches the Cu(001) surface along the $z$ direction. The probability corresponding to this process can be calculated as follows [53]:

$$P_{C\perp} = \int_{E_{b\perp}}^{\infty} p(E_k)dE_k = \left(1 - Erf[\sqrt{E_{b\perp}/(k_BT)}]\right) + \frac{2}{\sqrt{\pi}}\sqrt{E_{b\perp}/(k_BT)}e^{-\left(\frac{E_{b\perp}}{k_BT}\right)}. \quad (7)$$

As mentioned above, $p(E_k) = 2\pi(\frac{1}{\pi k_B T})^{\frac{3}{2}}\sqrt{E_k}e^{-E_k/(k_BT)}$ is the kinetic energy distribution function at a given temperature $T$, $E_k$ is the kinetic energy, and $E_{b\perp}$ corresponds to the height of $U(z)$, which is ~ 0.586 eV.

The second process is the dissociation due to the thermal motions of a single H atom, overcoming the constraint of H-H bond. The probability of this process is given by: $P_{C\parallel} = e^{-\left(\frac{E_{b\parallel}}{k_BT}\right)}$, where $E_{b\parallel} = E_{H-H}$ is the H-H bond energy (experimental data [74]: $E_{b\parallel}$ ~ 436 kJ/mole or equivalently, $E_{b\parallel}$ ~ 4.51 eV) . This value is well above $E_{b\perp}$ and thus one has $P_{C\perp} \gg P_{C\parallel}$. Therefore, dissociation in the horizontal direction is negligible compared to that in the vertical (surface normal) direction due to the translational motions of H$_2$. In summary, as a classical particle, the dissociation probability of H$_2$ is given by $P_c = P_{C\perp} + P_{C\parallel} \cong P_{C\perp}$. The calculated transmission probabilities are presented in Fig. 6, as a function of temperature. It is seen that both quantum and classical probabilities increase with elevating temperatures. Below $T =$ 1350 K, the transmission probability taking into account quantum tunneling effects ($P_Q$) is always higher than the case of treating the H$_2$ molecule as a classical particle ($P_C$). This phenomenon is even more pronounced at room temperature and below, at which the quantum transmission probability $P_Q$ is well above its classical counterpart $P_C$. The lower panels of Fig. 6 compare the contribution of the two processes. It is found that the first process plays a major role in the overall dissociation, especially at low temperatures. At $T \leq 18$ K, the probability of the first process is more than 100 orders of magnitude higher than that of the second one. In low energy region, the corner-cutting effects [75-79], due to which the tunneling pathway deviates from the MEP can be important for H$_2$ dissociation. Such effects have been partly included in the calculation of MEP, for which the constraint of symmetric configurations of H$_2$ is imposed (see texts above).

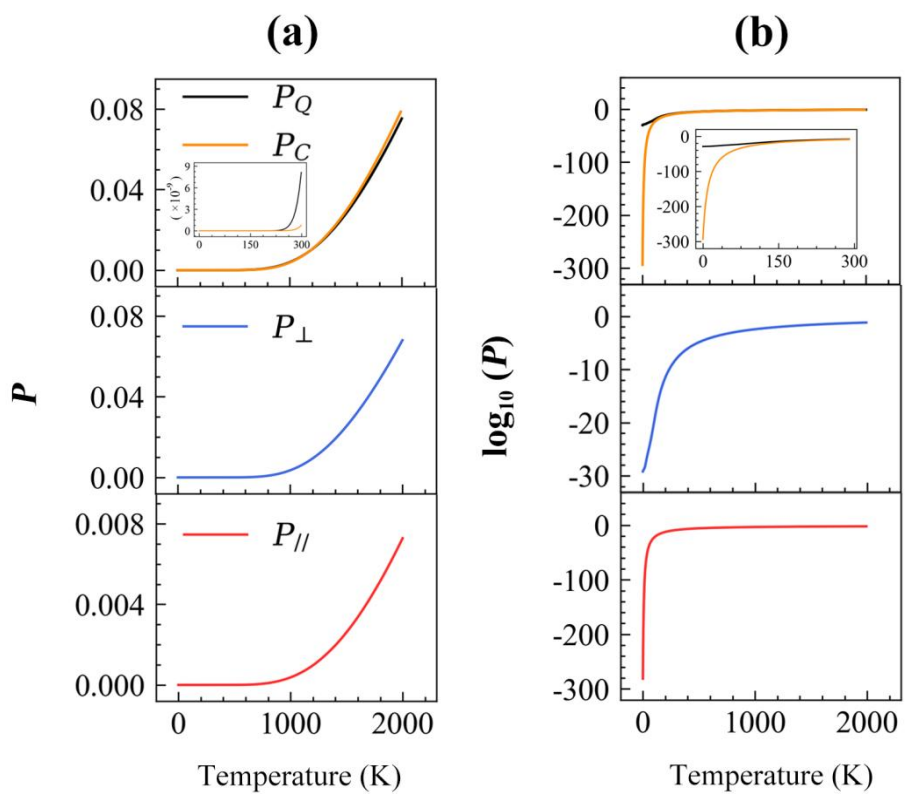

**Fig. 6**. The probability (panel (a)) and logarithms (panel (b)) of H$_2$ dissociation due to the translational motions in surface normal (vertical) direction (subscript ⊥) and vibrations in the transverse direction (subscript //) on Cu(001). The classical ($P_C$) and quantum probability ($P_Q$) are compared as a function of temperature.

Using the calculated dissociation probability of the H$_2$ molecule on Cu(001), we can estimate the rate constants of dissociation. The reaction rate constant $k$ can be obtained via the following relation [54]: $k = vP$, where $v$ is the attempting frequency and $P$ is the corresponding dissociation probability. For the first process, $v$ can be expressed as the reciprocal of the time $\tau(E)$ for travelling from the starting point $z_0$ to the end point $z_1$ (Fig. 5(a)). The dissociation rate constant for the first process is:

$$v_\perp(E) = \frac{1}{\tau(E)} = \frac{1}{\int_{z_0}^{z_1} \sqrt{\frac{m}{2(U(z)-E)}} dz}, \qquad (8)$$

where $E$ is the kinetic energy of the incident particle, $U(z)$ is the dissociation potential along the vertical direction as shown in Fig. 5(a), and $m$ is the mass of H$_2$ molecule. The traversal time $\tau(E)$ of tunneling across the barrier is obtained using the formula presented in Ref. [80]. The reaction rate constant of the first process is

$$k_\perp(T) = \int_0^\infty v_\perp(E) p(E,T) T_r(E) dE \cong \int_0^{E_m} v_\perp(E) p(E,T) T_r(E) dE. \qquad (9)$$

Like above, $E_m = 5E_b$ is adopted for numerical evaluation of the integral, and the kinetic energy distribution $p(E,T) = 2\pi(\frac{1}{\pi k_B T})^{3/2} \sqrt{E} e^{-E/k_B T}$, for temperature $T$.

For the second process, the term $v$ is the frequency of the H-H stretching mode in the transverse direction, $v_\parallel(z)$, which varies with height $z$. In fact, $v_\parallel(z)$ gets progressively smaller with decreasing height $z$. The rate constant for the second process is given by

$$k_\parallel(T) = \frac{1}{L}\int_{z_0}^{z_1} dz \int_{U(z)}^\infty v_\parallel(z) p(E,T) T_r(z) dE \cong \frac{1}{L}\int_{z_0}^{z_1} dz \int_{U(z)}^{E_m} v_\parallel(z) p(E,T) T_r(z) dE, \qquad (10)$$

where the vibrational frequency $v_\parallel(z)$ can be obtained from DFPT calculations. Ultimately, the total rate constant is

$$k_Q = k_\perp(T) + k_\parallel(T). \qquad (11)$$

The results are shown in Fig. 7. It is seen from the reaction rate constant that dissociation in the vertical direction dominates the whole process. On the other hand, it is seen that (insets of Fig. 7) dissociation due to H-H vibrations (horizontal stretching) plays a nontrivial role at $T \geq 300$ K, for which both the translational and vibrational motions of H$_2$ contribute significantly to the breaking of H-H bond.

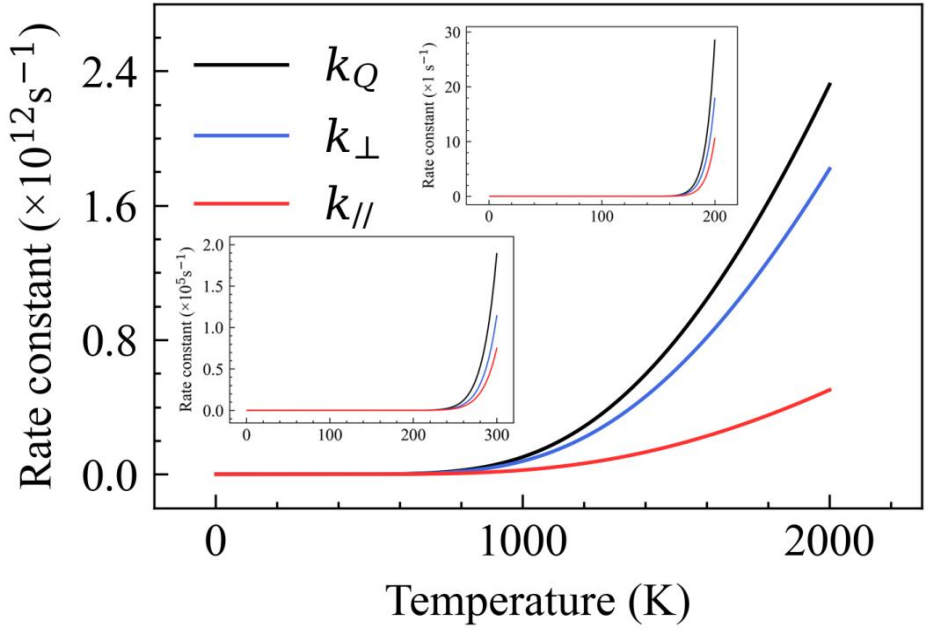

**Fig. 7.** Temperature dependence of quantum rate constant $k_Q$, and their vertical and transverse components.

At a temperature $T_d = 220$ K (insets of Fig. 7), the rate constant of dissociation is $k_Q \sim 300$/s, which corresponds to a dissociation number of $N_{H2} \approx 1.08 \times 10^6$ within one hour. The smallest face-centered two-dimensional (2D) surface unit cell of Cu(001) is a square with the side length of $a \sim 3.61$ Å and surface area $S_{Cu(001)} = a^2 \cong 13.03$ Å$^2$. Periodic extension by 1000 times along the two basis vectors gives a square substrate with a side length of $L \sim 0.361$ μm, which is comparable to size of samples employed experimentally. On average, each 2D surface unit cell can adsorb $1.08 \times 2$ H atoms. Define that adsorption of one H to one surface Cu gives a surface coverage of one monolayer (ML); then at $T_d = 220$ K, the surface coverage within one hour on Cu(001) is estimated to be $\sim 1.13$ ML. This is experimentally measurable, and nontrivial dissociation of H$_2$ is expected to take place on Cu(001) at $T \geq T_d$. The predicted temperature for the onset of nontrivial dissociation of H$_2$, $T_d = 220$ K, is comparable with previous experimental observations of thermal desorption of dissociated H from Cu(001) ($T_d \sim 218$ K) [15]. Significant increase in the reaction rate constant is found when the temperature reaches 1000 K and above (Fig. 7), in which dissociation due to the stretching mode of H-H vibrations starts to play a nontrivial role. This is also in good agreement with the previous experimental measurements. When the kinetic energies of the hydrogen molecules ejected from the molecular beam nozzle reached about an equivalent temperature of ~ 1000 K and above, it was observed that hydrogen began to adsorb instantaneously on the copper surfaces with measurable sticking coefficients [6].

The kinetic processes studied above are based on the precondition that the dosed $H_2$ molecules are in thermal equilibrium with well-defined kinetic energy distribution [54]. The situation changes significantly when the incident $H_2$ molecules are produced by molecular beam nozzle [6, 10], in which the kinetic energy of $H_2$ molecules distributes within a narrow range [10] and can be approximately taken as single-valued. For the monoenergetic molecular beams of $H_2$ with kinetic energy $E$, the probability of tunneling is given by the transmission coefficient $T_r(E)$, which is readily computed using the TM method. For a $H_2$ molecule which penetrates the barrier $U(z)$ and arrives at the Cu(001), the rate constant of dissociation $k_{mbz}(E) = v_\perp(E) \times P_\perp(E) = v_\perp(E) \times T_r(E)$. We have made comparison on the rate constant of dissociation of $H_2$ under thermal equilibrium and the non-equilibrium molecular beam experiment. The characteristic temperature for the activation of rotation degree of freedom of a $H_2$ molecule is $\theta_r$ ~ 85 K [81]. By contrast, the vibrational degree of freedom of $H_2$ is largely frozen at its ground state due to the very high characteristic temperature ($\theta_v$ ~ 6100 K) [81]. For temperatures above $\theta_r$, both the translational and rotational degree of freedom of $H_2$ are activated, and the equivalent temperature of the incident $H_2$ molecular beam with kinetic energy $E$ can therefore be estimated to $T = \frac{2E}{5k_B}$. The calculated rate constants are shown in Fig. 8, as a function of temperature and the kinetic energy of each $H_2$ from molecular beams. It is clearly seen that at the same temperature the rate constant under thermal equilibrium ($k_Q$) is larger than that of molecular beams ($k_{mbz}$). Compared to monoenergetic molecular beam with the translational kinetic energy $E_t$, the $H_2$ molecules at thermal equilibrium have a nontrivial probability of being at the energy states $E \geq E_t$, which is $P(E \geq E_t)$ $= \int_{E_t}^{\infty} p(E)dE \cong \frac{2}{\sqrt{\pi}}\sqrt{E_t/(k_BT)}e^{-(\frac{E_t}{k_BT})}$. Recalling that $E_t = \frac{3k_BT}{2}$, we have $P(E \geq E_t)$ $\cong \frac{2}{\sqrt{\pi}}\sqrt{3/2}e^{-(\frac{3}{2})} \cong 0.308$. This means that a considerable portion of $H_2$ molecules possess kinetic energies above the average $E_t$, and explains the difference between $k_Q$ and $k_{mbz}$.

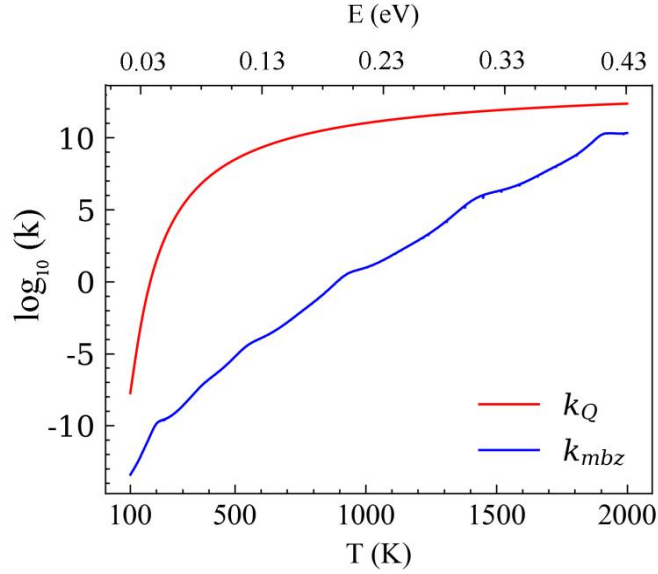

**Fig. 8**. Calculated rate constants $k$ (in logarithms) for the dissociation of $H_2$ on Cu(001), under the situation of thermal equilibrium ($k_Q$) and molecular beam experiments ($k_{mbz}$).

Using the experimental parameters presented in Ref. [10], i.e., the flux of the incident $H_2$ molecular beam through the nozzle (with diameter $d$ = 3 mm) is ~ $10^{15}$/sec, the number of incident $H_2$ molecules per second for unit area is estimated to be $F_{mbz} = \frac{N_{H2}}{S_{mbz}} = \frac{N_{H2}}{\pi(\frac{d}{2})^2} = \frac{10^{15}}{\pi \times 2.25 \times 10^{14}} \cong 1.41$ /Å$^2$. Within a time $t$, the number of $H_2$ molecules deposited on the surface unit cell of Cu(001) is therefore

$$n_{d,H_2}(E) = F_{mbz} \times S_{\text{Cu}(001)} \times k_{mbz}(E) \times t \qquad (12)$$

In the case $t$ = 1 sec, $n_{d,H_2}(E) \cong 18.4 \times k_{mbz}(E)$. Our calculations give that, for the kinetic energy $E$ = 0.176 eV, $k_{mbz}(E) \cong 0.073$/sec, and $n_{d,H_2}(0.176) \cong 1.3$, which implies that the surface coverage of dissociatively adsorbed H is ~ 1 ML. This is expected to give detectable signals for experimental measurements. The threshold value of $E$ = 0.176 eV for measurable dissociative adsorption of $H_2$ on Cu(001) is comparable with the experimental data ($E$ = 0.2 eV) based on molecular beam measurements [6]. The result points to the key role of quantum tunneling in the activated dissociation of $H_2$ on Cu(001).

### 3.3. The Effective barrier and isotope effects

For the barrier-crossing process of a microscopic particle that satisfies the Van't Hoff-Arrhenius relation, the total transmission probability ($P_{tot}$) can be generally expressed as follows [54]:

$$P_{tot}(T) = e^{-E_b^*/(k_B T)}, \qquad (13)$$

where $E_b^*$ is the effective barrier, and $k_B$, $T$ have the usual meanings as before. Therefore, the value of the effective barrier $E_b^*$ can be obtained as follows:

$$E_b^* = -(k_B T)\ln[P_{tot}(T)] = k_B T \ln[\frac{1}{P_{tot}(T)}]. \qquad (14)$$

Clearly, the effective barrier $E_b^*$ is a function of temperature. Using the probability $P_{tot}(T)$ obtained by the TM method, the values of $E_b^*$ at different temperatures are calculated and shown in Fig. 9. The data for the dissociation of $H_2$ and $D_2$ are presented, to show the isotope effects. It is seen that the effective barrier $E_b^*$ which taking into account the effects of quantum tunneling is appreciably lower than the classical barrier height (dissociation along the surface normal, $U(z)$) which is deduced by treating the incident $H_2$ molecules as classical particles. For $T \geq 200$ K, the effective dissociation barriers of both $H_2$ and $D_2$ increase smoothly with $T$ and arrive at their maxima at ~ 500 K, then decrease slowly with further increasing temperatures. The nearly steady gap ($\Delta E = E_{bH2}^* - E_{bD2}^* \sim 0.02$ eV) between the effective barriers of $H_2$ and $D_2$ is a consequence of quantum tunneling, whose probability decreases with increasing particle mass. Our calculations found that, for $T \geq 1000$ K the gap ($\Delta E$) decreases gradually and vanishes at ~ 3500 K, beyond which the system approaches the classical limit. The existence of a maximum of $E_b^*$ can be qualitatively understood based on Eq. (14), where the competition between the term $k_BT$ and $\ln[\frac{1}{P_{tot}(T)}]$ arrives at a compromise [54]. The weak temperature dependence of $E_b^*$ at $T \geq 300$ K indicates that it is usually sufficient to use one barrier parameter to describe the temperature-dependent kinetic process of atoms.

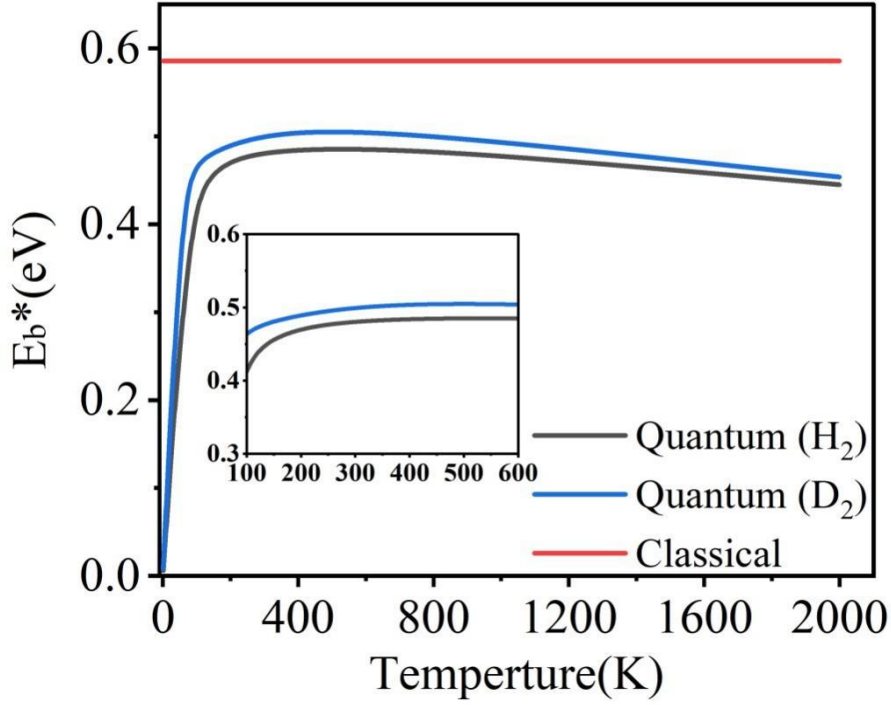

**Fig. 9**. The energy barriers for the dissociation of $H_2$ and $D_2$ on Cu(001). The original barrier height calculated using adiabatic approximation is referred to as "Classical". The effective barriers ($E_b^*$) due to quantum effects are presented as a function of temperature.

From $T \sim 200$ K to 260 K, the effective barrier for $H_2$ ($D_2$) increases slightly from $E_b^* \sim 0.47$ (0.49) eV to $E_b^* \sim 0.48$ (0.50) eV. This is comparable with the data from thermal desorption (reverse process of dissociation) measurements in the same range of temperature, in which the activation barrier of dissociation is determined to be $48 \pm 6$ kJ/mol ($\sim 0.50 \pm 0.06$ eV) for $H_2$ and $56 \pm 8$ kJ/mol ($0.58 \pm 0.08$ eV) for $D_2$, respectively. The situation is much different for $T \leq 200$ K. The effective barrier drops almost linearly with decreasing temperature. From 200 K to 10 K, the effective barrier for $H_2$ ($D_2$) decreases quickly from $E_b^* \sim 0.47$ (0.49) eV to $E_b^* \sim 0.06$ (0.08) eV.

When the temperature $T$ approaches 0 K, $E_b^*$ tends to be zero, which reveals that quantum tunneling effects may have a significant impact on the atomic scale dynamics even at extremely low temperatures.

The significant role of quantum tunneling can be further demonstrated by investigating the isotope effects on the transmission probability and rate constant of H$_2$ and D$_2$ dissociation on Cu(001). In calculations based on the TM method, both H$_2$ and D$_2$ are treated as quantum particles, for a temperature range of $T \leq 300$ K. The results are shown in Fig. 10. In the low temperature region, in particular, $T \leq 100$ K, significant differences between the transmission probabilities and rate constants are found. Using the relation $k = \nu P$, the natural logarithm of reaction constant, $\ln(K_Q)$ may be expressed as follows

$$\ln(K_Q) = \ln(\nu P_Q) = \ln\left(\nu e^{-\frac{E_b^*}{k_B T}}\right) = \ln\nu - \frac{E_b^*}{k_B T}, \quad (15)$$

where the minus slope of the $\ln(K_Q) \sim \frac{1}{k_B T}$ lines simply corresponds to the effective barrier $E_b^*$ at a given temperature $T$. As seen from Fig. 10(d), the turning point at $T \sim 100$ K distinguishes the rate constants of H$_2$ from D$_2$, an indication of crossover from classical to quantum regime. Moreover, the variation trends of the two curves can be roughly described by the combination of two lines with the slopes being equivalent to two effective barriers, as schematically shown by the dash lines in Fig. 10(d). Similar characteristics of $T$-dependent rate constant are found in the case of H diffusion on Cu(001) and Pt(111) surfaces [82-84], and may be expected for the diffusion processes on the other metal surfaces. In phenomenological descriptions of the effects of quantum tunneling [69, 85], two sets of parameters for the activation barriers (one for classical and one for quantum motions) plus two prefactors are introduced which roughly account for the dynamics at high and low temperature regions. By contrast, our method produces the major features of the variations of rate constant with temperature in a logically natural and unified manner, and provides a promising way for including the effects of quantum tunneling in the chemical dynamics of many-body systems.

The transmission coefficients $T_r(E)$, the total transmission probability $P_{tot}(T)$ at given temperature $T$, the rate constants $k$, and the corresponding effective barrier $E_b^*$ are deduced based on the PES calculated at the DFT-GGA level. When a more accurate PES for the H$_2$-Cu(001) interactions (e.g. the PES calculated by the semi-empirical SRP-DFT approach [42-45]) is employed, or when the van der Waals corrections are introduced, or when the interactions due to surface phonons are included [44]: All of which may be taken as perturbations to the PES obtained by DFT-GGA calculations. As has been shown [54] that perturbations/small changes to the original PES would only result in minor modifications on the transmission coefficients $T_r(E)$, and therefore the quantities derived based on $T_r(E)$. Consequently, the main results regarding the role of quantum tunneling in the dissociation kinetics of H$_2$ on Cu(001) are qualitatively kept unchanged.

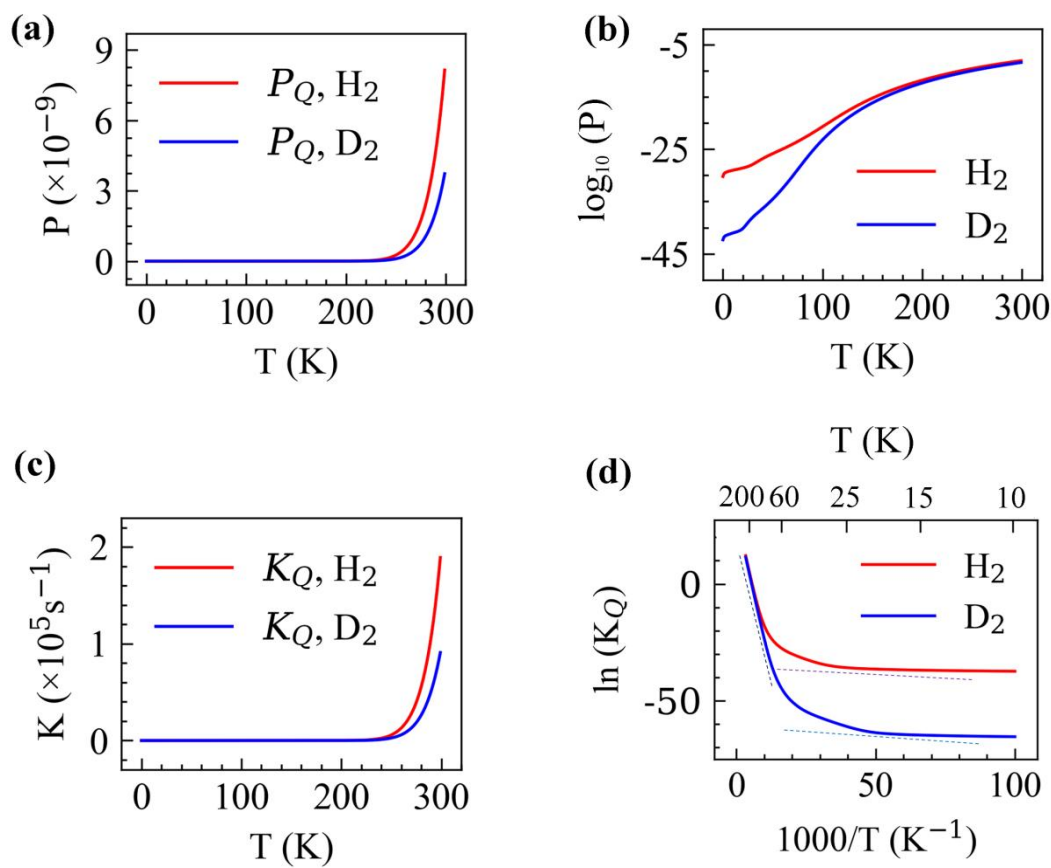

Fig. 10. Calculated quantum probabilities (upper panels) and rate constants (lower panels) for H$_2$ and D$_2$ dissociation on Cu(001), as function of temperature. The crossover from classical to quantum regime is schematically indicated by dash lines

(panel (d)).

## 4. Concluding Remarks

In summary, the activation and dissociation processes of $H_2$ molecules on Cu(001) surface have been studied based on first-principles calculations and the transfer matrix (TM) method. It is found that charge transfer from the substrate Cu atoms to $H_2$ plays an essential role in activating and breaking the H-H bond. Near the critical point of bond breaking, the adsorption configurations are characterized by the existence of two stationary states, and abrupt change of H-H bond length is expected upon small perturbations to the translational motions of $H_2$ as whole unit. Using the MEP determined by DFT calculations, the probabilities of $H_2$ molecules passing through the dissociation energy pathways are calculated using the TM method, which explicitly takes into account the effects of quantum tunneling. The probabilities and rate constants of dissociation due to the translational and vibrational motions are evaluated and distinguished. Both translational and vibrational motions contribute nontrivially to the dissociation of $H_2$ at high temperatures. For the situation where the Van't Hoff-Arrhenius relationship applies, the effective potential for the dissociation and adsorption of $H_2$ on Cu(001) is calculated. After considering the effects of quantum tunneling, the barrier height is significantly reduced with comparison to that of the barrier which treats the $H_2$ as a classical particle within the Born-Oppenheimer approximation. In studies on the situation of thermal equilibrium and the non-equilibrium molecular beam of $H_2$, the calculated temperatures for the onset of measurable dissociation of $H_2$ on Cu(001) are in agreement with experiments. To further demonstrate the role of quantum tunneling, we have computed the effective barriers of dissociation for both $H_2$ and $D_2$ which are found to be comparable with experimental data. The role of quantum effects is found to be remarkable at low-temperature region in which the crossover from classical to quantum regime is identified. The results are expected to be tested by future experimental works.

**Acknowledgements**

This work is financially supported by the National Natural Science Foundation of China (No. 11474285, 12074382). We are grateful to the staff of the Hefei Branch of Supercomputing Center of Chinese Academy of Sciences, and the Hefei Advanced Computing Center for support of supercomputing facilities. We also thank the referees for their reading and helpful comments.